\documentclass[reprint,preprintnumbers,nofootinbib,superscriptaddress,amsmath,amssymb,aps,prl]{revtex4-2}
\pdfoutput=1
\usepackage{graphicx,hyperref,braket,orcidlink, algorithm2e, amssymb, amsmath, dsfont, lineno, bm, slashed, multirow}
\usepackage[nolist,nohyperlinks]{acronym}
\usepackage{soul}
\usepackage{booktabs}
\usepackage{dcolumn}
\graphicspath{{figs/}}
\usepackage{tikz}
\usetikzlibrary{positioning,decorations.pathreplacing,decorations.markings,calc,arrows.meta}
\hypersetup{
	bookmarks=true,
	unicode=true,
	pdftoolbar=true,
	pdfmenubar=true,
	pdffitwindow=false,
	pdfstartview={FitH},
	pdftitle={pdf},
	pdfauthor={J.Y.~Araz, name-name}, 
	pdfsubject={pdf},
	pdfcreator={J.Y.~Araz, name-name},
	pdfproducer={}, 
	pdfkeywords={},
	pdfnewwindow=true,
	colorlinks=true,
	linkcolor=blue,
	citecolor=olive,
	filecolor=magenta,
	urlcolor=cyan
}

\definecolor{cgreen}{HTML}{2CA02C}
\definecolor{cblue}{HTML}{1F77B4}
\definecolor{corange}{HTML}{FF7F0E}
\definecolor{cyellow}{HTML}{FFD700}
\definecolor{cgrey}{HTML}{7F7F7F}
\definecolor{cred}{HTML}{D62728}
\definecolor{cpurple}{HTML}{9467BD}

\definecolor{org}{HTML}{FF4F00}

\newcommand{\psibar}{\bar{\psi}}
\newcommand{\Mh}{M_{h}}

\begin{document}

\title{Parton structure of mesonic states with tensor networks}

\author{Jack Y. Araz\orcidlink{0000-0001-8721-8042}}
\email{j.araz@ucl.ac.uk}
\affiliation{Department of Physics and Astronomy, University College London, London, WC1E 6BT, UK}
\affiliation{Department of Engineering, City St.\ George's, University of London, London, EC1V 0HB, UK}

\author{Yamil Cahuana Medrano\orcidlink{0009-0006-0492-0885}}
\email{yacahuanamedra@wm.edu}
\affiliation{Department of Physics, William and Mary, Williamsburg, VA 23185, USA}

\author{Peter L. Nguyen\orcidlink{0009-0004-9010-9321}}
\email{peter.nguyen@duke.edu}
\affiliation{Department of Physics, Duke University, Durham, NC 27708, USA}

\author{Zhong-Bo Kang\orcidlink{0000-0003-3161-0381}}
\email{zkang@physics.ucla.edu}
\affiliation{Department of Physics and Astronomy, University of California, Los Angeles, CA 90095, USA}
\affiliation{Mani L. Bhaumik Institute for Theoretical Physics, University of California, Los Angeles, CA 90095, USA}
\affiliation{Center for Quantum Science and Engineering, University of California, Los Angeles, CA 90095, USA}

\author{Jian-Wei Qiu\orcidlink{0000-0002-7306-3307}}
\email{jqiu@jlab.org}
\affiliation{Theory Center, Thomas Jefferson National Accelerator Facility, Newport News, VA 23606, USA}

\author{David Richards\orcidlink{0000-0001-6971-873X}}
\email{dgr@jlab.org}
\affiliation{Theory Center, Thomas Jefferson National Accelerator Facility, Newport News, VA 23606, USA}

\author{Nobuo Sato\orcidlink{0000-0002-1535-6208}}
\email{nsato@jlab.org}
\affiliation{Theory Center, Thomas Jefferson National Accelerator Facility, Newport News, VA 23606, USA}
\author{Marco Zaccheddu\orcidlink{0000-0002-3672-8111}}
\email{mzaccheddu@anl.gov}
\affiliation{Theory Center, Thomas Jefferson National Accelerator Facility, Newport News, VA 23606, USA}

\author{Jia-Yue Zhang\orcidlink{0000-0001-5594-4816}}
\email{jzhang@jlab.org}
\affiliation{Theory Center, Thomas Jefferson National Accelerator Facility, Newport News, VA 23606, USA}


\preprint{JLAB-THY-26-5052} 

\begin{abstract}
We present the first direct tensor-network calculation of flavor-resolved parton distribution functions (PDFs) for a pion-like isospin triplet in the two-flavor Nambu–Jona-Lasinio model in $(1+1)$ dimensions. Four-fermion interactions bind the mesons without confinement. Using staggered fermions and a charge-symmetrized Hamiltonian, we prepare the bound states and compute light-cone correlators by real-time evolution. The resulting quark and antiquark PDFs obey the expected isospin relations under $u\leftrightarrow d$ exchange between charged pions and exhibit equal $u$ and $d$ distributions in the neutral pion. A flavor-independent antisymmetrization of the field ordering removes the vacuum-subtracted anticommutator contribution and restores charge-conjugation relations in both the correlators and PDFs across the triplet within numerical accuracy. These results provide benchmarks for quantum simulations of the partonic structure of the meson.
\end{abstract}

\maketitle
\raggedbottom

\paragraph{Introduction}
Understanding how the quark and gluon degrees of freedom of quantum chromodynamics give rise to the observed properties of hadrons is a central goal of nuclear and particle physics. Parton distribution functions (PDFs) provide a fundamental characterization of this structure, describing the light-cone momentum fractions carried by partons and entering the factorized description of high-energy scattering processes~\cite{Collins:1989gx, Collins:2011zzd}. Determining PDFs from first principles remains challenging because they are defined through bilocal correlations along the light cone, whereas conventional lattice-QCD calculations are formulated in Euclidean time.

A variety of approaches relate quantities computed on a Euclidean lattice to parton structure through equal-time correlations of parton fields or currents~\cite{Ji:2013dva,PhysRevD.96.034025,PhysRevD.96.094503,Ma:2014jla, Alexandrou:2018pbm,Ma:2017pxb}. Other observables, such as fragmentation functions, require intrinsically time-dependent information~\cite{Collins:1981uw, Li:2024nod}. A Hamiltonian formulation provides access to real-time evolution and, hence, to light-cone correlations directly~\cite{Lamm:2019uyc, Echevarria:2020wct}. In low-dimensional theories, tensor networks provide a powerful implementation of this approach, enabling accurate preparation of interacting states and controlled real-time evolution~\cite{PhysRevB.73.094423, Haegeman:2015ezw}.

In recent years, there has been a growing number of Hamiltonian calculations of partonic observables. The QuNu collaboration has developed and benchmarked quantum algorithms for PDFs, light-cone distribution amplitudes, and generalized parton distributions~\cite{Li:2021kcs, Li:2022lyt, Li:2026thq}. Tensor-network calculations include PDFs and distribution amplitudes in the Nambu--Jona-Lasinio (NJL) model~\cite{Kang:2025xpz} and direct light-cone PDFs in the Schwinger model~\cite{Schneider:2024yub, Banuls:2025wiq}. Complementary studies of boosted equal-time quasi-distributions~\cite{Grieninger:2024cdl} now include tensor-network calculations of quasi-PDFs and quasi-GPDs~\cite{Grieninger:2025mbm}. Related efforts address fragmentation~\cite{Li:2024nod,Grieninger:2024axp}, real-time current correlators and hadronic tensors~\cite{Ikeda:2025bjb,Zou:2026cfk}, and PDFs on quantum hardware~\cite{Chen:2025zeh}.

In this Letter, we extend single-flavor tensor-network studies of PDFs~\cite{Kang:2025xpz, Banuls:2025wiq} by directly computing, for the first time, flavor-resolved PDFs for the full pion-like isotriplet in the two-flavor NJL model in $(1+1)$ dimensions~\cite{Nambu:1961tp, Nambu:1961fr}, which is also known as the chiral Gross-Neveu model with isospin~\cite{Thies:2022kuv}. This renormalizable and asymptotically free theory~\cite{Gross:1974jv, Boehmer:2009sw} binds the triplet without confinement. With two flavors, the scalar and independent isovector pseudoscalar interactions together preserve $SU(2)_L\times SU(2)_R$ chiral symmetry in the massless limit, providing distinct charged and neutral meson states.  We use real-time evolution to evaluate the light-cone correlators that define the PDFs.  The charged pions have valence content $u\bar d$ and $d\bar u$, whose nonzero net flavor numbers provide normalization conditions for the $q(x)-\bar q(x)$ distribution of quark flavor $q$. Isospin and charge conjugation also relate the shapes of the $q(x)$ and $\bar{q}(x)$ distributions within and across the triplet, providing symmetry tests unavailable in a single-flavor calculation~\cite{Shastry:2022obb}.

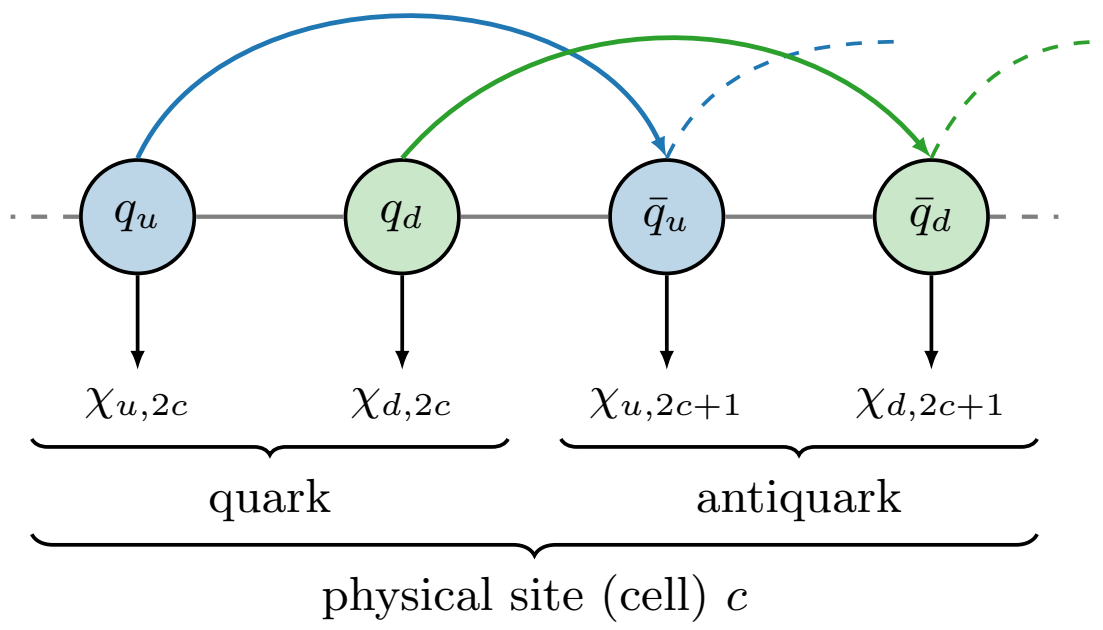
\begin{figure}[!tb]
    \centering
    \begin{tikzpicture}[
            scale=1.4, line width=0.7pt, font=\small,
            site/.style={draw, circle, minimum size=7.5mm, inner sep=0pt},
            uq/.style={site, fill=cblue!30},
            dq/.style={site, fill=cgreen!25},
            lab/.style={font=\small},
            hop/.style={line width=0.9pt, -{Latex[length=1.4mm]}}]
          \def\dx{1.25}
          \foreach \i/\x/\st/\nm in {0/0/uq/{q_{u}}, 1/1/dq/{q_{d}}, 2/2/uq/{\bar q_{u}}, 3/3/dq/{\bar q_{d}}}{
            \node[\st] (s\i) at (\x*\dx,0) {$\nm$};
          }
          \foreach \a/\b in {0/1,1/2,2/3}{\draw[thick,cgrey] (s\a)--(s\b);}
          \draw[thick,cgrey,dashed] (s3)--++(0.6,0);
          \draw[thick,cgrey,dashed] (s0)--++(-0.6,0);
          \foreach \i/\n in {0/{\chi_{u,2c}}, 1/{\chi_{d,2c}}, 2/{\chi_{u,2c+1}}, 3/{\chi_{d,2c+1}}}{
            \draw[-{Latex[length=1.3mm]}] (s\i.south)--++(0,-0.45);
            \node[lab,anchor=north] at ($(s\i.south)+(0,-0.45)$) {$\n$};
          }
          \draw[hop,cblue] (s0.north) to[out=65,in=115] (s2.north);
          \draw[cblue,dashed] (s2.north) to[out=65,in=180] ++(1.1,0.55);
          \draw[hop,cgreen] (s1.north) to[out=50,in=130] (s3.north);
          \draw[cgreen,dashed] (s3.north) to[out=65,in=180] ++(0.8,0.55);
          \def\yb{-1.05}
          \draw[decorate,decoration={brace,mirror,amplitude=3pt}] ({-0.4*\dx},\yb) -- ({1.4*\dx},\yb)
            node[midway,below=4pt,lab]{quark};
          \draw[decorate,decoration={brace,mirror,amplitude=3pt}] ({1.6*\dx},\yb) -- ({3.4*\dx},\yb)
            node[midway,below=4pt,lab]{antiquark};
          \draw[decorate,decoration={brace,mirror,amplitude=4pt}] ({-0.4*\dx},-1.5) -- ({3.4*\dx},-1.5)
            node[midway,below=4pt,lab]{physical site (cell) $c$};
    \end{tikzpicture}
    \caption{Lattice layout of one physical site (cell $c$) of the tensor network. Circles are the four staggered sites (i.e.\ MPS tensors), with their physical legs below; color marks the flavor ($u$ blue, $d$ green), and the quark and antiquark rows are labeled. Colored arrows show the nearest-neighbor hops in Eq.~\eqref{njl:eq:H}. Each flavor forms its own staggered chain on alternating qubits, so an intra-flavor hop spans two MPS bonds.}
    \label{fig:lattice}
\end{figure}

\paragraph{Hamiltonian formulation} For an isospin doublet $\psi=(u,d)^T$ with degenerate bare mass $m$, the continuum NJL Hamiltonian density is
\begin{align}
\mathcal{H} ={}&\psibar\big(-i\gamma^{1}\partial_x + m\big)\psi\nonumber\\
&-\frac{g}{2}\Big[(\psibar\psi)^{2} + \sum_{k=1}^{3}\big(\psibar\,i\gamma_{5}\tau_{k}\,\psi\big)^{2}\Big],
\label{eq:continuum_hamiltonian}
\end{align}
where $g$ is the dimensionless four-fermion coupling, $\psibar=~\psi^{\dagger}\gamma^0$, and $\vec\tau$ are the Pauli matrices in flavor space. For one flavor, a Fierz identity relates the pseudoscalar interaction to the scalar term, making it redundant. We discretize the fields using staggered fermions~\cite{Kogut:1974ag, Susskind:1976jm}, with $\psi_q(x_c)=a^{-1/2}(\chi_{q,2c},\chi_{q,2c+1})^T$ for $q\in\{u,d\}$ and cells $c=0,\dots,N_c-1$ at $x_c=2ca$. Here $a$ is the staggered-site spacing. Each cell contains four fermion modes, represented by four qubits. The lattice Hamiltonian is
\begin{align}
\label{njl:eq:H}
H ={}& -\frac{i}{2a}\sum_{q=u,d}\sum_{n=0}^{2N_c-2} \Big(\chi_{q,n}^{\dagger}\chi_{q,n+1}-\mathrm{h.c.}\Big) \nonumber\\
 &+ \sum_{q=u,d} m\sum_{n=0}^{2N_c-1}(-1)^{n}\,\chi_{q,n}^{\dagger}\chi_{q,n} \nonumber\\
 &- \frac{g}{a}\sum_{c=0}^{N_c-1}\Big[S_{c}^{2} + \sum_{k=1}^{3} \big(P^{k}_{c}\big)^{2}\Big],
\end{align}
where $N_c$ denotes the number of cells. The scalar-isoscalar density $S_c$ and pseudoscalar-isovector densities $P_c^k$ ($k=1,2,3$) are the staggered transcriptions of $\bar\psi\psi$ and $\bar\psi i\gamma_5\tau_k\psi$. As shown in Fig.~\ref{fig:lattice}, we order the four fermion modes in each cell as $\{\chi_{u,2c},\chi_{d,2c},\chi_{u,2c+1},\chi_{d,2c+1}\}$, with subscripts denoting flavor and staggered site. The explicit bilinears are given in \hyperref[app:lattice]{Appendix A}.

For degenerate masses $m_u=m_d=m$, exact isospin symmetry, $[H,I^2]=[H,I_3]=0$, allows states to be classified by $(I,I_3)$. Charge conjugation obeys $\{C,I_3\}=0$ and $[C,I^2]=0$, mapping $(I,I_3)$ to $(I,-I_3)$. In the staggered Hamiltonian, charge conjugation preserves the hopping and mass terms but exchanges the interaction pairings $(2c,2c+1)$ and $(2c+1,2c+2)$. We therefore define the lattice charge-conjugation operator~\cite{Golterman:1984cy, Catterall:2024jps, Chatterjee:2024gje}
\[
\mathcal{C}: \chi_{q,n} \longmapsto \chi^{\dagger}_{q,n+1}
\]
and define the charge-symmetrized Hamiltonian through
\begin{equation}
    H_C = \frac{1}{2} \left(H + \mathcal{C} H \mathcal{C}^\dagger\right).
\end{equation}
With periodic boundary conditions, $\mathcal{C}$ exchanges the two cell pairings, so $[H_C,\mathcal{C}]=0$. With open boundary conditions, the shifted cell at the edge is absent, leaving a boundary contribution to the commutator. In a naive expansion in the lattice spacing, the difference between $H$ and $H_C$ is suppressed by one power of $a$, so they have the same classical continuum limit~\cite{Symanzik:1983dc}.

\hyperref[app:lattice]{Appendix A} explains why the first-order pion energy shift between $H$ and $H_C$ vanishes for periodic boundaries.

\paragraph{State preparation and light-cone correlators} We use ITensors.jl~\cite{Fishman:2020gel} and DMRG~\cite{White:1992zz} with bond dimension capped at $2000$ to prepare the vacuum and the lowest-energy pion-like meson states.  
The net quark number $Q=N_q-N_{\bar q}$ counts quarks minus antiquarks across both flavors, relative to the filled Dirac sea. In the $Q=0$ sector, the vacuum is the $I=0$ ground state of the charge-symmetrized Hamiltonian $H_C$, and $\pi^+$, $\pi^0$, and $\pi^-$ are the lowest-energy states with $I=1$ and $I_3=+1,0,-1$, respectively. 
We use $N_c=51$ unit cells ($204$ qubits), $g=0.08$, and $m_u=m_d=0.1$
in lattice units ($a=1$), leaving $m_u\neq m_d$ isospin breaking situation for future study. We use open boundaries because, at a given numerical accuracy, they generally require smaller matrix-product-state bond dimensions than periodic boundaries~\cite{Verstraete:2004zza}. These pion states are standing waves rather than exact momentum eigenstates; we use them as finite-volume approximations to pions at rest.

\begin{table}[!t]
\caption{Classification of hadronic states with total quark number $Q=0$.  The first column denotes the usual spectroscopic identification; ``content'' denotes the valence quark content, and the following two columns are isospin.  The final column shows the excitation energies above the vacuum in lattice units.  For reference, we also include the lowest state with $Q=1$, which corresponds to a constituent quark.\label{njl:tab:sectors}}
\begin{ruledtabular}
\begin{tabular}{llccc}
state             & content                    & $2I_3$ & $2I$   & $M_X$\\ \hline\hline
vacuum            & $-$                        & $0$    & $0$   & $-$\\
quark             & $u$                        & $+1$   & $1$ & 0.3883 \\
\midrule
$\pi^{+}$         & $u\bar d$                  & $+2$   & $2$   & 0.5924\\
$\pi^{-}$         & $d\bar u$                  & $-2$   & $2$   & 0.5924 \\
\midrule
$\pi^{0}$         & $(u\bar u-d\bar d)/\sqrt2$ & $0$    & $2$ & 0.5925\\
\end{tabular}
\end{ruledtabular}
\end{table}

The excitation energies above the vacuum, $M_X=~E_X-E_\Omega$ for $X\in\{q,\pi^\pm,\pi^0\}$, are listed in Table~\ref{njl:tab:sectors}. The independently optimized pion states are nearly degenerate, as expected from isospin symmetry. In lattice units, the charged-pion mass difference is $|M_{\pi^+}-M_{\pi^-}|\simeq2\times10^{-12}$, while the neutral--charged difference is $(M_{\pi^0}-M_{\pi^+})\simeq1.1\times10^{-4}$. The latter is small compared with the neutral-pion energy width, $\sigma_{E,\pi^0}=\sqrt{\langle H_C^2\rangle_{\pi^0}-\langle H_C\rangle_{\pi^0}^2}=1.2\times10^{-2}$, which measures the residual departure of the prepared state from an exact energy eigenstate. The same residual limits the isospin relation $C^{\pi^0}_{q}=(C^{\pi^+}_{u}+C^{\pi^+}_{d})/2$ to $3.4\%$ of the correlator peak. The pions are bound, $M_\pi<2M_q$, while the Hamiltonian also admits quark-like eigenstates, reflecting the absence of confinement.

\begin{figure*}[!th]
    \centering
    \includegraphics[width=0.9\linewidth]{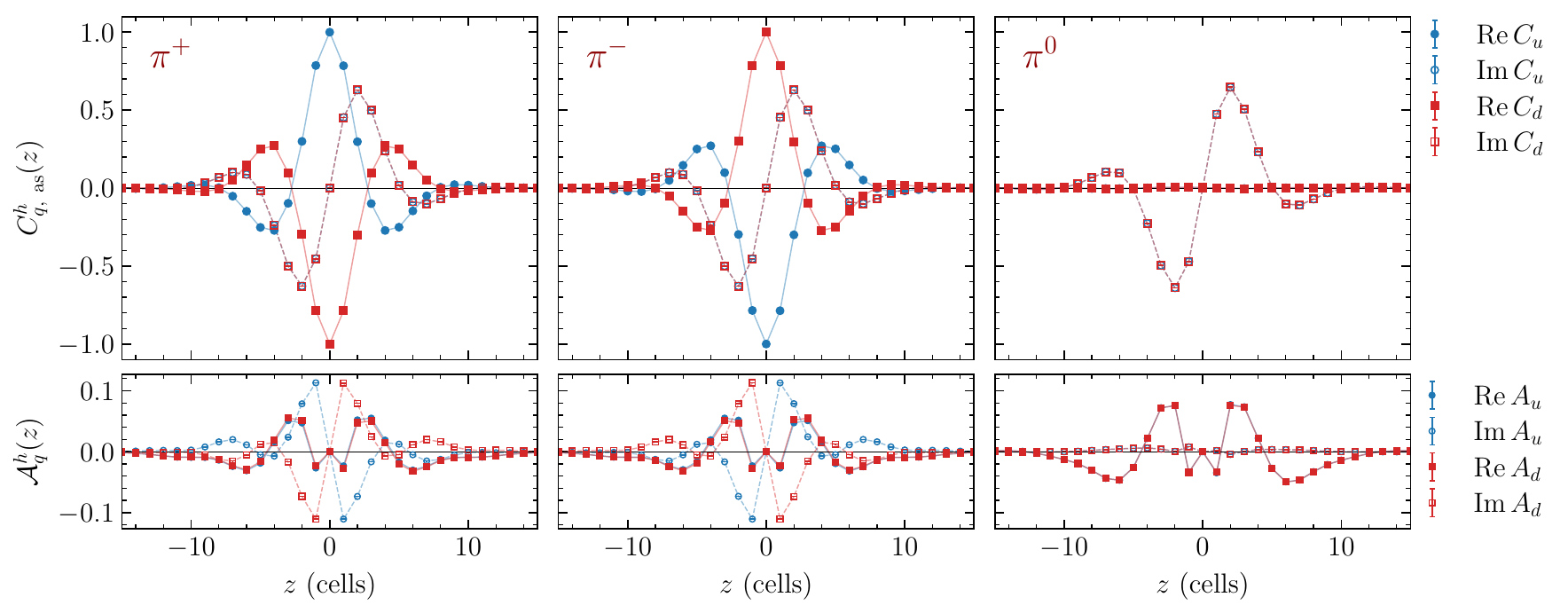}
    \caption{Connected pion matrix elements versus separation $z$ in cells. The top row shows the correlator antisymmetrized in field order, $C_{q,\mathrm{as}}^h$, and the bottom row shows the vacuum-subtracted anticommutator $\mathcal{A}_q^h$ for the same field pair. Both rows include the normalization by $N_c/\mathcal R$, so the top row is $\mathcal M(\nu_z)$ of Eq.~\eqref{njl:eq:itd}. Columns show $\pi^+$, $\pi^-$, and $\pi^0$ from left to right. Blue circles and red squares denote $u$ and $d$; filled and open symbols denote real and imaginary parts, respectively.}
    \label{fig:spatial_me}
\end{figure*}

In the NJL model, the bare, vacuum-subtracted PDF $f_{q/h}(x)$ describing the light-cone momentum fraction $x$ carried by a \emph{bare quark} $q$ in a hadron $h$ is given by
\begin{eqnarray}
    f_{q/h}(x) &=& \int_{-\infty}^{\infty} \frac{dz}{4\pi} e^{-izx\, n \cdot P} \langle h({P})|\mathcal{O} | h({P}) \rangle_{\mathrm{sub}} \nonumber\\
    &\stackrel{P^1=0}{=}& \int_{-\infty}^{\infty} \frac{dz}{4\pi} e^{-izxM_h}\langle h|\mathcal{O}|h\rangle_{\mathrm{sub}}\ , \label{eq:pdf}
\end{eqnarray}
where $\mathcal{O}=\bar\psi_q(z {n})\, {n}\cdot {\gamma}\,\psi_q({0})$, $\psi_q$ is the quark field of flavor $q\in\{u,d\}$, ${P}$ is the hadron energy--momentum vector, and ${n}=(1,-1)$ is lightlike. The subscript $\mathrm{sub}$ denotes the matrix element of $\mathcal{O}-\langle\Omega|\mathcal{O}|\Omega\rangle$, with $|\Omega\rangle$ the vacuum state. The PDFs reported here are bare, with no operator renormalization applied. Since continuum PDFs are invariant under longitudinal boosts, we follow the rest-frame construction of Ref.~\cite{Li:2021kcs} and evaluate the second line at ${P}=(M_h,0)$, with $|h\rangle\equiv|h(M_h,0)\rangle$.

We antisymmetrize the two field orderings using
\begin{equation}
\mathcal O_{\mathrm{as}}(z)=\frac12\Big[\mathcal O-\operatorname{tr}_{D}\!\big[\psi_q(0)\bar\psi_q(z n)\, n\cdot\gamma\big]\Big].
\label{eq:antisym_operator}
\end{equation}
The trace contracts spinor indices without changing the order of the displayed fields. We use vacuum-subtracted matrix elements of this operator for both flavors throughout the pion triplet. In the continuum, the projected fields lie at equal light-front time, although their ordinary times differ; their anticommutator is a state-independent contact term that cancels upon vacuum subtraction (Sec.~6.9.3 of Ref.~\cite{Collins:2011zzd}). Thus $\langle h|\mathcal O_{\mathrm{as}}|h\rangle_{\mathrm{sub}}=\langle h|\mathcal O|h\rangle_{\mathrm{sub}}$, preserving the continuum PDF definition. This equality fails for our discrete lattice correlators, as discussed below.

In our discrete Hamiltonian theory, we first define the ordered operator at arbitrary time and endpoint cells,
\begin{equation}
\label{njl:eq:Mz}
\hat{\mathcal O}_q(t;c_1,c_2)=\sum_{i,j\in\{0,1\}}
\chi^\dagger_{q,2c_1+i}(t)\,\chi_{q,2c_2+j}(0),
\end{equation}
where $c_1$ and $c_2$ label the sink and source cells, and $i,j$ select their two staggered spinor components. The vacuum-subtracted correlator is $C_q^h(t;c_1,c_2)=~\langle h|\hat{\mathcal O}_q(t;c_1,c_2)|h\rangle_{\mathrm{sub}}$. Applying Eq.~\eqref{eq:antisym_operator} at the same endpoints defines $\hat{\mathcal O}_{q,\mathrm{as}}$ and $C_{q,\mathrm{as}}^h$.

For the PDF calculation, we fix the source at the central cell $c_0$ and sample the light cone with $t=2az$ and $c_1=c_0-z$. We write $\hat{\mathcal O}(z)\equiv\hat{\mathcal O}_q(2az;c_0-z,c_0)$ and likewise for $\hat{\mathcal O}_{\mathrm{as}}$.

Converting unit-normalized DMRG states to the continuum normalization supplies the factor $2M_hL$, with $L=2N_ca$ (\hyperref[app:lattice]{Appendix A}). To account for the enhanced central density of open-boundary standing waves, we use $\mathcal R=N_c C_q^h(0;c_0,c_0)/\sum_c C_q^h(0;c,c)\approx1.85$. These coincident-endpoint correlators are unscaled; for vanishing spatial current, their sum is the net flavor number, $\pm1$ for the charged pions. For the $\pi^0$, whose net flavor numbers vanish, we use the charged-pion value of $\mathcal R$. Henceforth, $C_q^h(z)\equiv C_q^h(2az;c_0-z,c_0)/\mathcal R$, and likewise for $C_{q,\mathrm{as}}^h(z)$. Equal-time vacuum subtraction gives $C_{q,\mathrm{as}}^h(0)=C_q^h(0)$, so $N_c C_{q,\mathrm{as}}^h(0)=\pm1$ for charged pions. This imposes the valence normalization rather than testing it independently.

To impose the translation-invariant conjugation relation in the open box, we average $C_{q,\mathrm{as}}^h(z)$ with $[C_{q,\mathrm{as}}^h(-z)]^*$. The PDFs of the approximate rest states then follow from the lattice Fourier transform~\cite{Kang:2025xpz, Banuls:2025wiq}
\begin{equation}
\label{njl:eq:pdf}
f_{q/h}(x) = \frac{2\Mh L}{4\pi}\sum_{z} e^{-i x \Mh (2za)}\, C_{q,\mathrm{as}}^h(z).
\end{equation}

To evaluate these light-cone matrix elements, we apply the fermion operators at the source and sink sites to the prepared state $|h\rangle$, forming matrix-product states $\chi_{q,2c_0+j}|h\rangle$ and $\chi_{q,2(c_0-z)+i}|h\rangle$. We compute the unequal-time matrix elements from overlaps involving the time-evolved field-inserted states, including the phase associated with the energy of $|h\rangle$. We repeat the calculation for the vacuum subtraction and use creation-operator insertions for the reversed field ordering. We perform the state evolution with the time-dependent variational principle (TDVP)~\cite{Haegeman:2011zz, Haegeman:2015ezw}, using bond-dimension caps $D=600$ and $800$, truncation cutoff $10^{-10}$, time step $a/2$, and separations $|z|\le z_{\max}=25$. We find no discernible difference between the results obtained with $D=600$ and $800$. All results are provided with an estimated truncation error for $D=600$ to $800$, which is on the order of $\mathcal{O}(10^{-4})$.

\paragraph{Symmetry tests of the correlators} We test the correlators using isospin and charge conjugation, which are exact symmetries of the continuum theory. Charge conjugation interchanges quarks with their antiquarks. Combined with the isospin rotation that interchanges $u$ and $d$, it maps each charged pion onto itself and exchanges $u\leftrightarrow\bar d$ and $d\leftrightarrow\bar u$, so that
\begin{equation}
u^{\pi^+}(x)=\bar d^{\pi^+}(x),\qquad \bar u^{\pi^+}(x)=d^{\pi^+}(x),\label{njl:eq:cpred}
\end{equation}
and likewise $d^{\pi^-}=\bar u^{\pi^-}$, $\bar d^{\pi^-}=u^{\pi^-}$. With two flavors, these combined symmetries thus relate partons of different flavors inside a single hadron. In a one-flavor theory, charge conjugation only implies that the neutral meson, its own charge conjugate, has $q=\bar q$, so its valence distribution vanishes identically, and neither a separation into valence and sea nor a flavor relation can be tested.

The upper row of Fig.~\ref{fig:spatial_me} shows that a single, flavor-independent antisymmetrization recovers the charge-conjugation and isospin relations across the pion triplet, within the resolution shown. For the $\pi^+$, the $u$ and $d$ correlators have opposite real parts and nearly identical imaginary parts, $C_{u,\mathrm{as}}^{\pi^+}\simeq-[C_{d,\mathrm{as}}^{\pi^+}]^*$, to $0.3\%$ of the peak. Isospin interchanges $u$ and $d$ between the charged pions. For the $\pi^0$, the two flavor correlators coincide, $C_{u,\mathrm{as}}^{\pi^0}\simeq C_{d,\mathrm{as}}^{\pi^0}$, and their real parts are consistent with zero, $\mathrm{Re}\,C_{q,\mathrm{as}}^{\pi^0}\simeq0$, below $0.8\%$ of the imaginary peak.

The effect of this prescription is shown in the lower row of Fig.~\ref{fig:spatial_me}, where the change in the correlators is quantified by
\begin{align}
\mathcal A_q^h(z)
&=\frac{2}{\mathcal R}\langle h|\hat{\mathcal O}(z)-\hat{\mathcal O}_{\mathrm{as}}(z)|h\rangle_{\mathrm{sub}}\nonumber\\
&=2\big[C_q^h(z)-C_{q,\mathrm{as}}^h(z)\big].
\label{eq:anticommutator_residue}
\end{align}
At $z=0$, the canonical anticommutator is state independent, so vacuum subtraction gives $\mathcal A_q^h(0)=0$. At nonzero separations, all three pion states exhibit an oscillatory signal that decreases at larger separations. The charged-pion flavor contributions have nearly equal real parts and opposite imaginary parts, while the neutral-pion contributions coincide and are predominantly real. For the $\pi^0$, the original correlator has a real part reaching about $7\%$ of its peak imaginary part; subtracting $\mathrm{Re}\,\mathcal A_q^{\pi^0}/2$ yields the near-zero real part shown in the upper row.

The nonzero residue is consistent with interacting unequal-time evolution producing a state-dependent anticommutator, whose pion and vacuum expectation values need not cancel. Equal-time canonical relations remain intact; in the free theory, linear evolution makes the anticommutator state independent. Although the projected continuum fields obey a canceling contact-term relation at equal light-front time, spatial discretization does not guarantee this cancellation for our samples. The staggered-site offsets and an exploratory Wilson-fermion check are discussed in \hyperref[app:lattice]{Appendix A}. The residue is stable over the different volumes studied, so increasing lattice size at fixed spacing does not remove it. Its lattice-spacing dependence and continuum limit remain for future work.

For $x>0$, quark and antiquark distributions are identified through $q(x)=f_{q/h}(x)$ and $\bar q(x)=-f_{q/h}(-x)$ (Sec.~6.9.3 of Ref.~\cite{Collins:2011zzd}). Our antisymmetrization prescription enforces the field-ordering relation required for this identification by removing the vacuum-subtracted anticommutator contribution.

\paragraph{PDF reconstruction} Finite correlator samples leave a null space of PDF variations invisible to the sampled Fourier transform, making reconstruction an ill-posed inverse problem~\cite{Dutrieux:2024rem}. We use Gaussian process regression (GPR) to infer distributions and quantify their uncertainty~\cite{Medrano:2025cmg, Candido:2024hjt}. Directions unconstrained by the data remain prior-dependent.

To formulate the reconstruction, we express Eq.~\eqref{njl:eq:pdf} in terms of the sampled Ioffe times $\nu_z=2zaM_h$ as
\begin{equation}
\mathcal{M}(\nu_z)\equiv\frac{L}{2a}\,C_{q,\mathrm{as}}^h(z)=\int_{-x_{\mathrm{BZ}}}^{x_{\mathrm{BZ}}}dx\,e^{ix\nu_z}f_{q/h}(x).
\label{njl:eq:itd}
\end{equation}
Here $z$ is an integer cell separation. At fixed $a$ and $M_h$, $\Delta x=2az\,n$ and $P=(M_h,0)$ give the scalar Ioffe time $\nu_z=P\cdot\Delta x=2azM_h$. The sampling interval $\Delta\nu=~2aM_h$ sets $x_{\mathrm{BZ}}=\pi/(2aM_h)\simeq2.65$. We reconstruct over this full zone, allowing finite-spacing support beyond $|x|=1$, with $\mathcal M(0)$ fixing its valence integral.







The correlators cover separations $-25\le z\le25$. After imposing the conjugation relation between opposite separations, the GPR fits use the nonnegative Ioffe times up to $\nu_{\max}=2z_{\max}aM_\pi\approx30$. The field-ordering antisymmetrization enforces the relation needed to identify $\bar q(x)=-f_{q/h}(-x)$ for $x>0$. Splitting the Fourier integral in Eq.~\eqref{njl:eq:itd} into positive and negative $x$ then gives
\begin{align}
\operatorname{Re}\mathcal M(\nu)
&=\int_0^{x_{\mathrm{BZ}}}dx\,\cos(x\nu)\,f^-_{q/h}(x),\nonumber\\
\operatorname{Im}\mathcal M(\nu)
&=\int_0^{x_{\mathrm{BZ}}}dx\,\sin(x\nu)\,f^+_{q/h}(x),
\label{eq:gp_channels}
\end{align}
where $f^-_{q/h}=q-\bar q$ and $f^+_{q/h}=q+\bar q$. For each hadron and flavor, we apply GPR independently to the real and imaginary parts of $\mathcal M$ to reconstruct these two combinations. The real-part fit uses $26$ values, including $z=0$, and the imaginary-part fit uses $25$ values with $z>0$.

We reconstruct the PDFs using quadratic finite elements and independent, zero-mean GPR priors with a covariance smooth in $\ln x$. The prior variance, correlation length, and common Gaussian residual scale are determined by maximizing the marginal likelihood~\cite{Rasmussen:2006gpml}. We report posterior means and pointwise $68\%$ credible bands, conditional on these fitted hyperparameters. The discretization, prior, likelihood, and posterior are specified in the \hyperref[app:gp]{Appendix B}.

\paragraph{Flavor-resolved parton distributions} We obtain quark and antiquark PDFs from $q=(f^++f^-)/2$ and $\bar q=~(f^+-f^-)/2$. For the $\pi^+=u\bar d$, the valence combinations are $u_v=u-\bar u=f^-_{u/\pi^+}$ and $\bar d_v=\bar d-d=-f^-_{d/\pi^+}$, while $\bar u$ and $d$ are the sea distributions. The $\pi^-$ follows by $u\leftrightarrow d$.

\begin{figure}[!t]
    \centering
    \includegraphics[width=0.95\linewidth]{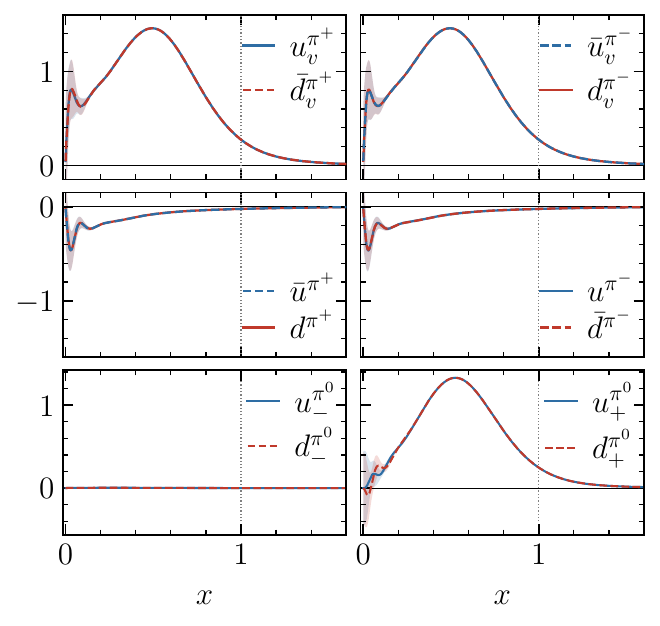}
    \caption{Reconstructed pion PDFs from antisymmetrized correlators, normalized with $\mathcal R$. The top and middle rows show charged-pion valence and sea distributions, respectively, with $\pi^+$ on the left and $\pi^-$ on the right. The bottom row shows the neutral-pion combinations $f^-_{q/\pi^0}=q-\bar q$ (left) and $f^+_{q/\pi^0}=q+\bar q$ (right). Blue and red denote the $u$ and $d$ flavors, respectively; the legends identify the quark and antiquark distributions. Curves and bands show GPR posterior means and pointwise $68\%$ credible intervals. Vertical dotted lines mark $x=1$; the reconstruction extends to $x_{\mathrm{BZ}}\simeq2.65$.}
    \label{fig:valsea}
\end{figure}

In Fig.~\ref{fig:valsea}, we plot the reconstructed PDFs obtained from the antisymmetrized correlators. The distributions extend beyond the continuum boundary $x=1$ and decay toward zero as $x$ approaches the Brillouin-zone edge $x_{\mathrm{BZ}}\simeq2.65$; the displayed range emphasizes their dominant structure. The upper two rows show a clear realization of the combined charge-conjugation and isospin relations. The paired valence and sea curves overlap at the resolution shown, and interchanging $u$ and $d$ maps the $\pi^+$ onto the $\pi^-$. This agreement follows only after applying the same field-ordering antisymmetrization to every flavor, as discussed above. In the bottom row, the neutral-pion flavor distributions agree within the reconstruction uncertainties and $f^-_{q/\pi^0}$ is consistent with zero on the scale shown, realizing $q^{\pi^0}=\bar q^{\pi^0}$ for each flavor.

The charged-pion valence distributions are positive and broad, peaking near $x=0.5$, while the sea is negative and smaller in magnitude. This hierarchy resembles the positive valence and suppressed non-valence contributions discussed in Ref.~\cite{Collins:2021vke}, which show that renormalized PDFs need not be positive. Their renormalization mechanism does not, however, explain the sign of our bare lattice PDFs; antisymmetrization restores symmetry relations, not positivity. The sign is a property of the correlators, not of the reconstruction or of the field ordering. Positivity is guaranteed for light-front number densities, because every parton carries $k^+>0$~\cite{Brodsky:1997de, Collins:2021vke}. A vacuum-subtracted density of bare lattice quanta has no such bound. To estimate the bare pair content, we expand each flavor field in the eigenmodes $a_k$ of the free lattice Hamiltonian with mass $m$ and count antiquarks as vacancies of the negative-energy modes, $\bar N=\sum_{\epsilon_k<0}\langle a_k a_k^\dagger\rangle$, evaluated at equal time from the one-body density matrix $\langle\chi^\dagger_{q,n}\chi_{q,n'}\rangle$ of each state. We find that the interacting vacuum contains approximately three bare $q\bar q$ pairs per flavor, and that the $\pi^+$ contains approximately $7\%$ fewer $\bar u$ antiquarks. This depletion permits a negative sea. Understanding the origin and continuum behavior of the sea will help clarify the partonic interpretation of these lattice distributions.

The GPR bands widen at small $x$, where the finite Ioffe-time range increases prior sensitivity, with characteristic resolution $\Delta x\sim1/\nu_{\max}$~\cite{Dutrieux:2024rem}. Larger lattices, together with controlled evolution to longer times, would extend the reliable Ioffe-time range and improve these constraints.

\paragraph{Conclusions}  

We present the first direct tensor-network calculation of flavor-resolved light-cone PDFs for the pion-like triplet of the two-flavor NJL model, recovering its isospin structure. An interaction-dependent, vacuum-subtracted anticommutator contribution obstructs the usual field-ordering relation at unequal times, although equal-time canonical relations remain intact. A common, flavor-independent antisymmetrization removes this contribution and restores the expected correlator and PDF symmetry relations within the numerical accuracy resolved here. Its continuum behavior remains to be established. These results provide new benchmarks for quantum simulations of real-time partonic dynamics and further study of the partonic interpretation of lattice distributions.

\paragraph{Acknowledgments} We used Claude Opus 5.5 and ChatGPT in this project to search the parameter space, parallelize the benchmarks on the cluster, and check calculations. The authors checked all results and take full responsibility for the content. This work is supported by the U.S.\ Department of Energy (DoE), Office of Science, Office of Nuclear Physics under Contract No.~89243126CSC000213.  The research described in this paper was conducted in part under the Laboratory-Directed Research and Development Program at Thomas Jefferson National Accelerator Facility for the DoE.  The research reported in this work made use of computing facilities of the USQCD Collaboration, which are funded by the Office of Science of the DoE. JYA is supported by the Institute for Particle Physics Phenomenology Associateship Scheme and by the Royal Society under grant no. IES/R2/252139. ZK is supported by the National Science Foundation under grant No.~PHY2515057. YC acknowledges support in part from the Southeastern Universities Research Association (SURA) through Grant C2024-FEMT-011-03, awarded by the Center for Nuclear Femtography (CNF) and in part by the Fulbright-Garcia Robles scholarship. We are grateful for helpful conversations with Joe Karpie and Felix Ringer. 

\bibliography{bibliography}

\section*{End Matter}
\appendix*
\paragraph{Appendix A: Lattice implementation and checks}
\label{app:lattice}
The interleaved ordering in Fig.~\ref{fig:lattice} keeps the interaction between flavors within each cell local along the matrix-product-state chain, while nearest-neighbor hopping of a given flavor spans two bonds.

With $\gamma^0=\sigma^z$, $\gamma^1=i\sigma^y$, and $\gamma_5=\sigma^x$, the bilinears in Eq.~\eqref{njl:eq:H} are
\begin{align*}
S_c &= \sum_{q=u,d}\Big(\chi_{q,2c}^{\dagger}\chi_{q,2c}
      -\chi_{q,2c+1}^{\dagger}\chi_{q,2c+1}\Big),\\
P_c^k &= i\sum_{q,q'=u,d}(\tau_k)_{qq'}\Big(
         \chi_{q,2c}^{\dagger}\chi_{q',2c+1}
         -\chi_{q,2c+1}^{\dagger}\chi_{q',2c}\Big).
\end{align*}
Here the Pauli matrices $\sigma^i$ act on spinor components, while $\tau_k$ acts on flavor.

Wilson fermions preserve charge conjugation exactly, motivating their use in recent Schwinger-model GPD calculations~\cite{Li:2026thq}, but the Wilson term explicitly breaks chiral symmetry~\cite{Karsten:1980wd}. Staggered fermions instead admit a remnant discrete chiral symmetry associated with a one-site translation~\cite{Dempsey:2022nys}.

For periodic boundaries, the perturbation $\delta H=H-H_C$ changes sign under $\mathcal{C}$. Its expectation value therefore vanishes in any eigenstate of $\mathcal{C}$. The charged pions are not individually $\mathcal{C}$ eigenstates, but isospin makes their expectation values of $\delta H$ equal, whereas charge conjugation makes them opposite. Both expectation values must therefore vanish. Since the first-order energy shift is precisely this expectation value, the leading shift for an isolated level is second order in $\delta H$, formally $\mathcal{O}(a^2)$ in the naive lattice-spacing expansion, where $\delta H=\mathcal O(a)$. Open-boundary terms can spoil this cancellation, so this periodic-boundary argument does not eliminate boundary effects in our calculation.

The two-site DMRG algorithm used for state preparation optimizes two neighboring matrix-product-state tensors at each step.

Our DMRG states obey $\langle h|h\rangle=1$, whereas the continuum PDF definition uses $\langle P|P'\rangle=4\pi E_P\delta(P-P')$. The finite-volume identification $2\pi\delta(0)\to L$, with $L=2N_c a$, gives the rest-frame conversion factor $2M_hL$.

In the free lattice theory, time evolution is linear in the fermion fields, so the unequal-time anticommutator is proportional to the identity and cancels under vacuum subtraction. Four-fermion interactions introduce products of fields into their evolution, allowing a state-dependent anticommutator without modifying the equal-time canonical algebra.

For the staggered light-cone samples, the cell centers lie on the same light front, but the individual site pairs in Eq.~\eqref{njl:eq:Mz} have $\Delta t+\Delta x=(i-j)a$. Thus some component pairs are offset from the light front, and the discretized evolution does not guarantee the continuum cancellation. An exploratory exact-diagonalization check with Wilson fermions on a three-site periodic lattice also yields a nonzero vacuum-subtracted anticommutator at light-cone separation, despite exact charge-conjugation symmetry of the Hamiltonian and neutral-pion state. The residue vanishes to numerical precision at equal time and in the free theory. It is therefore not specific to the separation of spinor components across staggered sites.

\setcounter{equation}{0}
\renewcommand{\theequation}{B\arabic{equation}}
\renewcommand{\theHequation}{B\arabic{equation}}
\paragraph{Appendix B: Gaussian-process reconstruction}
\label{app:gp}
We implement the reconstruction with quadratic finite elements, writing $f^\pm_{q/h}(x)\simeq\sum_j w_j^\pm\phi_j(x)$, where $\phi_j$ are local interpolation functions on a positive-$x$ mesh extending to $x_{\mathrm{BZ}}$ and $w_j^\pm$ are the PDF values at its nodes. Integrating each basis function against the cosine or sine kernel gives the matrix $B$ that maps these coefficients to the sampled correlators. We assign independent, zero-mean GPR priors to the two distributions, with covariance
\begin{equation}
K(x,x')=s\exp\!\left[-\frac{(\ln x-\ln x')^2}{2\ell^2}\right].
\label{eq:gp_prior}
\end{equation}
Here $s$ sets the prior variance and $\ell$ controls the correlation length in $\ln x$. Larger $\ell$ favors variation over broader ranges of $x$, while the logarithmic coordinate permits finer structure at small $x$~\cite{Dutrieux:2024rem}.

For each channel, we model data as $\boldsymbol y=B\boldsymbol w+\boldsymbol\epsilon$, with zero-mean Gaussian residuals and covariance $\langle\epsilon_i\epsilon_j\rangle=\sigma^2\delta_{ij}$. The likelihood is
\begin{equation}
p(\boldsymbol y\mid\boldsymbol w,\sigma)
=\frac{\exp\!\left[-\frac{\|\boldsymbol y-B\boldsymbol w\|^2}{2\sigma^2}\right]}{(2\pi\sigma^2)^{N_{\mathrm{data}}/2}},
\label{eq:gp_likelihood}
\end{equation}
where $N_{\mathrm{data}}$ is the number of data points in the channel. Thus $\sigma$ is a common absolute residual scale across the sampled Ioffe times; it allows a mismatch between the computed correlators and their GPR reconstruction. We determine $\sigma$ jointly with $s$ and $\ell$ by numerically maximizing the GPR marginal likelihood, an empirical-Bayes procedure~\cite{Rasmussen:2006gpml}. The marginal likelihood integrates Eq.~\eqref{eq:gp_likelihood} over the PDF coefficients under their Gaussian prior, giving a zero-mean Gaussian for $\boldsymbol y$ with covariance $\Sigma=BKB^T+\sigma^2 I$, where $K_{jk}=K(x_j,x_k)$ is the prior covariance at the mesh nodes. With the fitted hyperparameters fixed, Bayes' theorem gives the Gaussian posterior
\begin{align}
p(\boldsymbol w\mid\boldsymbol y,s,\ell,\sigma)
&=\frac{p(\boldsymbol y\mid\boldsymbol w,\sigma)\,p(\boldsymbol w\mid s,\ell)}{p(\boldsymbol y\mid s,\ell,\sigma)}=\mathcal N(\boldsymbol w;\boldsymbol\mu,V)\nonumber\\
\boldsymbol\mu&=KB^T\Sigma^{-1}\boldsymbol y,\\
V&=K-KB^T\Sigma^{-1}BK.\nonumber
\label{eq:gp_posterior}
\end{align}
For either PDF combination, the posterior mean is $\bar f(x)=\boldsymbol\phi(x)^T\boldsymbol\mu$ and its variance is $\operatorname{Var}[f(x)]=\boldsymbol\phi(x)^T V\boldsymbol\phi(x)$, where $\boldsymbol\phi$ collects the finite-element basis functions. We report the mean and pointwise $68\%$ credible bands, conditional on the fitted hyperparameters.

\end{document}